\documentstyle[preprint,aps]{revtex}

\begin{document}
\draft
\title{Writing Spin in a Quantum Dot with Ferromagnetic and Superconducting
Electrodes}
\author{{\ Yu Zhu}$^1${, Qing-feng Sun}$^2${, and }Tsung-han Lin$^{1,*}$}
\address{$^1${\it State Key Laboratory for Mesoscopic Physics and }\\
{\it Department of Physics, Peking University,}{\small \ }\\
{\it Beijing 100871, China}}
\address{$^2${\it Center for the Physics of Materials and }\\
{\it Department of Physics, McGill University, }\\
{\it Montreal, PQ, Canada H3A 2T8}}
\maketitle

\begin{abstract}
We propose an efficient mechanism for the operation of writing spin in a
quantum dot, which is an ideal candidate for qubit. The idea is based on the
Andreev reflection induced spin polarization (ARISP) in a ferromagnetic /
quantum-dot / superconductor system. We find that on the resonance of
Andreev reflection, the spin polarization of quantum dot {\it strongly}
denpends on the magnetization of ferromagnetic electrode, and the {\it sign}
of the spin polarization is controllable by bias voltage. In the presence of
intradot Coulomb interaction, we show that ARISP effect can still survive as
long as the charging energy $U$ is comparable to the superconducting gap $%
\Delta $. Detailed conditions and properties of ARISP are also discussed.
\end{abstract}



\baselineskip 20pt 
\newpage

Qubit is the basic unit in the achievement of quantum computing. Among many
quantum two-level systems, the spin of quantum dot (QD) is an ideal
candidate for such purpose \cite{spin1}. Firstly, QD fabricated in
semiconductor 2DEG can be well controlled by metallic gates, parameters such
as resonant levels, tunnel barriers, and charging energy are experimentally
tunable. Second, the spin coherent time is extremely long in semiconductors
(exceeding 100ns at low temperature \cite{spin2}), which is orders larger
than charge coherent time. Moreover, due to 0D nature of QD, many spin-flip
mechanisms are further suppressed , resulting in an even longer lifetime of
spin in QD \cite{spin3}.

One of the challenge to exploit the spin of QD as qubit is to efficiently
control the local spin polarization in QD. A natural idea is to apply a
magnetic field and induce the spin polarization by Zeeman splitting \cite
{spin4}. But the scheme contains practical difficulties because the field is
required to be of the order of Tesla but confined within the scale of QD. An
alternative idea is to attach QD to ferromagnetic electrodes \cite
{spin4,spin5}. In principle, the asymmetry between two spin bands in the
electrodes may induce spin-dependent renormalization and broadening to the
resonant level of QD and lead to spin polarization. However, such mechanism
requires relatively large magnetization in electrodes and relatively strong
coupling between QD and electrodes. More important, the sign of spin
polarization in QD can not be controlled electrically.

It is the purpose of this paper to present an efficient mechanism for
writing spin in a QD with ferromagnetic (F) and superconducting (S)
electrodes. The idea is motivated by the fact that F can provide
spin-polarized electrons while S\ can only accept Cooper pair which is spin
singlet. We shall show below that on the resonance of Andreev reflection
(AR) \cite{AR1,AR2,AR3,AR4}, the spin polarization of QD strongly depends on
the magnetization of F, and the sign of spin polarization is controllable by
the bias voltage. The effect can survive even in the presence of strong
Coulomb interaction, as long as the charging energy of QD is comparable to
the superconducting gap in S.

{\it non-interacting case}--To see the physics clearly, we start with the
non-interacting ferromagnetic / quantum-dot / superconductor (F-QD-S) system
modelled by the following Hamiltonian, $H=H_F+H_S+H_{dot}+H_T$, in which $%
H_F=\sum_{k\sigma }(\varepsilon _k-\sigma h)a_{k\sigma }^{\dagger
}a_{k\sigma }$ is for the F electrode, $H_S=\sum_{p\sigma }\varepsilon
_pb_{p\sigma }^{\dagger }b_{p\sigma }+\sum_p(\Delta b_{p\uparrow }^{\dagger
}b_{-p\downarrow }^{\dagger }+H.c.)$ for the S electrode, $%
H_{dot}=\sum_\sigma E_\sigma c_\sigma ^{\dagger }c_\sigma $ is for the
noninteracting QD, and $H_T$ describes the tunnel couplings between QD and
electrodes. We introduce the retarded, advanced, and lesser Green function
in the Nambu representation: 
\begin{equation}
{\bf G}^{r,a,<}\equiv \int dt\;e^{\text{i}\omega t}\left( 
\begin{array}{cc}
\langle \langle c_{\uparrow }(t)|c_{\uparrow }^{\dagger }(0)\rangle \rangle
^{r,a,<} & \langle \langle c_{\uparrow }(t)|c_{\downarrow }(0)\rangle
\rangle ^{r,a,<} \\ 
\langle \langle c_{\downarrow }^{\dagger }(t)|c_{\uparrow }^{\dagger
}(0)\rangle \rangle ^{r,a,<} & \langle \langle c_{\downarrow }^{\dagger
}(t)|c_{\downarrow }(0)\rangle \rangle ^{r,a,<}
\end{array}
\right) \;.
\end{equation}
The occupation numbers $\langle n_{\uparrow }\rangle \equiv \langle
c_{\uparrow }^{\dagger }c_{\uparrow }\rangle $ and $\langle n_{\downarrow
}\rangle \equiv \langle c_{\downarrow }^{\dagger }c_{\downarrow }\rangle $
are related to$\;$the integral of ${\bf G}^{<}$.

In the non-interacting case considered here, ${\bf G}^r$, ${\bf G}^a$, and $%
{\bf G}^{<}$ can be solved exactly in Keldysh formalism as ${\bf G}^r{\bf =}%
\left( {\bf g}_0^{-1}-{\bf \Sigma }_0^r\right) ^{-1}$, ${\bf G}^a=\left( 
{\bf G}^r\right) ^{\dagger }$, and ${\bf G}^{<}={\bf G}^r{\bf \Sigma }_0^{<}%
{\bf G}^a$, where ${\bf g}_0$ is the Green function of the isolated QD (see
Eq.(7)) and ${\bf \Sigma }_0$ the self energy caused by the tunnel coupling
with electrodes. In the wide band limit, ${\bf \Sigma }_0$ can be evaluated
analytically as ${\bf \Sigma }_0{\bf =\Sigma }_{F0}{\bf +\Sigma }_{S0}$ with 
\begin{eqnarray}
\Sigma _{F0}^r &=&-\frac{\text{i}}2\left( 
\begin{array}{cc}
\Gamma _{F\uparrow } & 0 \\ 
0 & \Gamma _{F\downarrow }
\end{array}
\right) ,\;\Sigma _{F0}^{<}=\text{i}\left( 
\begin{array}{cc}
\Gamma _{F\uparrow }f(\omega -V) & 0 \\ 
0 & \Gamma _{F\downarrow }f(\omega +V)
\end{array}
\right) , \\
\Sigma _{S0}^r &=&-\frac{\text{i}}2\Gamma _S\rho ^r(\omega )\left( 
\begin{array}{cc}
1 & -\frac \Delta \omega \\ 
-\frac \Delta \omega & 1
\end{array}
\right) ,\;\Sigma _{S0}^{<}=\text{i}\Gamma _S\rho ^{<}(\omega )f(\omega
)\left( 
\begin{array}{cc}
1 & -\frac \Delta \omega \\ 
-\frac \Delta \omega & 1
\end{array}
\right) ,
\end{eqnarray}
in which $\Gamma _{F\uparrow }$, $\Gamma _{F\downarrow }$, and $\Gamma _S$
are the coupling strengths between QD and electrodes (notice that $\Gamma
_{F\uparrow }\neq \Gamma _{F\downarrow }$), $f(\omega )\equiv 1/(e^{\omega
/k_BT}+1)$ is the Fermi function, $V$ is the bias voltage between F and S 
\cite{remark1}, $\rho ^r(\omega )$ and $\rho ^{<}(\omega )$ are the
generalized BCS density of states defined as$\;\rho ^r(\omega )=\frac{\left|
\omega \right| }{\sqrt{\omega ^2-\Delta ^2}}\theta (\left| \omega \right|
-\Delta )+\frac \omega {\text{i}\sqrt{\Delta ^2-\omega ^2}}\theta (\Delta
-\left| \omega \right| )$ and $\rho ^{<}(\omega )=\frac{\left| \omega
\right| }{\sqrt{\omega ^2-\Delta ^2}}\theta (\left| \omega \right| -\Delta )$%
.

Fig.1 shows the curves of the occupation number $\langle n_\sigma \rangle $
vs the resonant level $E_0\equiv E_{\uparrow }=E_{\downarrow }$, with the
ratio $\Gamma _{F\downarrow }/\Gamma _{F\uparrow }=\frac 13$. (a), (b), and
(c) are corresponding to the bias voltage $V>0$, $V=0$, and $V<0$,
respectively. For $V=0$, there is a step from 0 to 1 in the occupation
curve, indicating an electron filling when $E_0$ passes the chemical
potential $\mu _F=\mu _S$. Moreover, the curves of $\langle n_{\uparrow
}\rangle $ and $\langle n_{\downarrow }\rangle $ are almost overlapped. For $%
V>0$ ($V<0$), the step is shifted to $E_0=\mu _F=V$, and there emerges a
resonant dip (peak) pinned at $E_0=\mu _S=0$. The results can be understood
as follows (see also the schematic diagram in the bottom): Since no single
particle states are available within the superconducting gap, electron
filling to the QD is mainly determined by the F electrode, resulting in a
step linked to the chemical potential of F. When the resonant level lines up
with the chemical potential of S, however, two-particle process AR may
occur, in which a spin$\uparrow $ and a spin$\downarrow $ electron in QD can
leak into S by forming a Cooper pair and vice versa \cite{Andreev}. As a
result, an Andreev dip (Andreev peak) is superposed on the step-like curve.
The most remarkable features of AR dip (AR peak) are: (1) The spin
polarization of QD, $m\equiv \langle n_{\uparrow }\rangle -\langle
n_{\downarrow }\rangle $, {\it strongly} depends on the magnetization of F.
(2) The {\it sign} of $m$ is controllable by the bias voltage, $m>0$ for $%
V>0 $ and $m<0$ for $V<0$. Hereafter, these nontrivial properties of AR
resonance is referred to as Andreev reflection induced spin polarization
(ARISP). Qualitatively, ARISP effect can be interpreted as follows: For $V>0$
and $E_0=\mu _S$, spin$\uparrow $ and spin$\downarrow $ electrons in QD tend
to form Cooper pair and enter S. Since F can provide more spin$\uparrow $
electron that spin$\downarrow $ electron, the spin$\downarrow $ electron
will be depleted by spin$\uparrow $ electron in the AR process, resulting in
a strong spin polarization in QD. For $V<0$ and $E_0=\mu _S$, a Cooper pair
is converted into a spin$\uparrow $ and a spin$\downarrow $ electron in QD.
It is much easier for spin$\uparrow $ electron than spin$\downarrow $
electron to escape to the empty states of F, resulting in a reversed spin
polarization.

For quantitative analysis, we evaluate $\langle n_\sigma \rangle $ near the
AR resonance and obtain $\langle n_{\uparrow }\rangle =1-n(p)$ and $\langle
n_{\downarrow }\rangle =1-n(-p)$ for $V>0$, $\langle n_{\uparrow }\rangle
=n(p)$ and $\langle n_{\downarrow }\rangle =n(-p)$ for $V<0$, where 
\begin{equation}
n(p)\equiv \frac 12(1-p)\frac{\Gamma _S^2}{4E_0^2(1-p^2)+\Gamma
_F^2(1-p^2)+\Gamma _S^2}\;,
\end{equation}
in which $p\equiv (\Gamma _{F\uparrow }-\Gamma _{F_{\downarrow }})/(\Gamma
_{F\uparrow }+\Gamma _{F\downarrow })$ is the magnetization in F and $\Gamma
_F\equiv (\Gamma _{F\uparrow }+\Gamma _{F\downarrow })/2$ is the spin
averaged coupling strength. Fig.2a shows the AR dip for different
magnetization $p$ of F electrode (also true for AR peak if upside down the
plot). Notice that the resonance has the Lorentzian line shape with
half-width $\sqrt{\left[ \Gamma _F^2(1-p^2)+\Gamma _S^2\right] /4(1-p^2)}$,
implying that AR resonance is broadened with the increase of $p$. Fig.2b
shows the maximum spin polarization $m_0=m(E_0=0)$ vs $p$ for different
coupling strength ratio $r\equiv \Gamma _F/\Gamma _S$. Contrary to the
intuition, ARISP effect is most pronounced when the ratio $r$ is small.
Especially, in the limit $r\rightarrow 0$, $\langle n_{\uparrow }\rangle _0=%
\frac 12(1\pm p)$ and $\langle n_{\downarrow }\rangle _0=\frac 12(1\mp p)$
(upper sign for $V>0$ and lower sign for $V<0$). This means that both F and
S electrodes are important in the ARISP effect: F provides the asymmetry
between two spin categories, while S reinforces the asymmetry through AR
process. Another noteworthy feature of ARISP is that the maximum spin
polarization is determined by the ratio $r$ and $p$, rather than the
magnitudes of $\Gamma _{F\uparrow }$, $\Gamma _{F\downarrow }$, and $\Gamma
_S$. Below we shall take into account the intradot Coulomb interaction to
investigate whether ARISP can still survive. Due to the symmetry between
electron and hole, only the case of $V>0$ is discussed.

{\it interacting case}--To include the Coulomb interaction, we add the term $%
Un_{\uparrow }n_{\downarrow }$ to $H_{dot}$, which makes F-QD-S a strong
correlated system. Notice that in the limit $U\rightarrow \infty $, double
occupation is forbidden in QD, and electron number can not fluctuate by two.
Hence AR is completely killed by Coulomb interaction in this limit. If,
however, the charging energy $U$ is comparable to the superconducting gap $%
\Delta $, AR can still occur with the aid of bias voltage. Therefore,
techniques for $U\rightarrow \infty $ limit (e.g., slave boson method) can
not be applied to the calculation of ARISP. Alternatively, we adopt the
equation of motion (EOM) method, which is known to be reliable in the
Coulomb blockade regime, and qualitatively correct for Kondo physics \cite
{EOM1,EOM2}. Moreover, EOM solution becomes exact in the $U\rightarrow 0$
limit. In the EOM approach, one can derive the equation for the retarded
Green function $\langle \langle A(t_1)|B(t_2)\rangle \rangle ^r$ by
differentiating with respect to $t_1$ or $t_2$, with new Green functions
generated in the equation. To close the equation chain, we make the
truncation in those Green functions containing two electrode operators in a
mean field manner. After some algebra, the equation for ${\bf G}^r$ can be
obtained as 
\begin{equation}
{\bf AG}^r+{\bf G}^r{\bf \tilde{A}=2N+B+\tilde{B}\;},
\end{equation}
in which ${\bf B\equiv (g}_1^{-1}{\bf -\Sigma }_1^r{\bf )U}^{-1}$, ${\bf %
A\equiv B(g}_0^{-1}{\bf -\Sigma }_0^r{\bf )+\Sigma }_2^r$, ${\bf \tilde{A}}$
or ${\bf \tilde{B}}$ represents the transpose of ${\bf A}$ or ${\bf B}$, $%
{\bf U}$ and ${\bf N}$ are defined as 
\begin{equation}
{\bf U\equiv }\left( 
\begin{array}{cc}
+U & 0 \\ 
0 & -U
\end{array}
\right) ,\;{\bf N\equiv }\left( 
\begin{array}{cc}
\langle n_{\downarrow }\rangle & 0 \\ 
0 & \langle n_{\uparrow }\rangle
\end{array}
\right) .
\end{equation}
${\bf g}_0$ and ${\bf g}_1$ are the bare Green function for the resonances $%
E_\sigma $ and $E_\sigma +U$, 
\begin{equation}
{\bf g}_0=\left( 
\begin{array}{cc}
\frac 1{\omega -E_{\uparrow }} & 0 \\ 
0 & \frac 1{\omega +E_{\downarrow }}
\end{array}
\right) ,\;{\bf g}_1=\left( 
\begin{array}{cc}
\frac 1{\omega -E_{\uparrow }-U} & 0 \\ 
0 & \frac 1{\omega +E_{\downarrow }+U}
\end{array}
\right) .
\end{equation}
${\bf \Sigma }_0^r$ and ${\bf \Sigma }_1^r$ are the corresponding dressing
self energies due to tunnel coupling with electrodes. ${\bf \Sigma }_2^r$ is
the self energy contributed by the spin flip in the cotunneling process,
which is related to Kondo physics. In the wide band limit, these self
energies can be evaluated analytically \cite{remark2}. Let ${\bf \Sigma }_i^r%
{\bf =\Sigma }_{Fi}^r{\bf +\Sigma }_{Si}^r$, (find ${\bf \Sigma }_{F0}^r$
and ${\bf \Sigma }_{S0}^r$ in Eq.(2) and Eq.(3)) 
\begin{eqnarray}
{\bf \Sigma }_{F1}^r &=&-\frac{\text{i}}2\left( 
\begin{array}{cc}
\Gamma _{F\uparrow }+2\Gamma _{F\downarrow } & 0 \\ 
0 & \Gamma _{F\downarrow }+2\Gamma _{F\uparrow }
\end{array}
\right) , \\
{\bf \Sigma }_{F2}^r &=&\left( 
\begin{array}{cc}
\Gamma _{F\downarrow }Q\left( \frac{-\omega _1-V}{k_BT},\frac{\omega _3-V}{%
k_BT}\right) & 0 \\ 
0 & \Gamma _{F\uparrow }Q\left( \frac{-\omega _3-V}{k_BT},\frac{\omega _2-V}{%
k_BT}\right)
\end{array}
\right) , \\
{\bf \Sigma }_{S1}^r &=&\Gamma _S\left( 
\begin{array}{cc}
s_2(\omega _0)+s_2(\omega _1)+s_2(\omega _3) & s_1(\omega _3) \\ 
s_1(\omega _3) & s_2(\omega _0)+s_2(\omega _2)+s_2(\omega _3)
\end{array}
\right) , \\
{\bf \Sigma }_{S2}^r &=&\frac{\Gamma _S}2\left( 
\begin{array}{cc}
s_2(\omega _1)+s_2(\omega _3)+s_4(\omega _1)-s_4(\omega _3) & s_1(\omega
_0)+s_1(\omega _3)+s_3(\omega _0)+s_3(\omega _3) \\ 
s_1(\omega _0)+s_1(\omega _3)-s_3(\omega _0)-s_3(\omega _3) & s_2(\omega
_2)+s_2(\omega _3)+s_4(\omega _3)-s_4(\omega _2)
\end{array}
\right) ,
\end{eqnarray}
in which $s_1(\omega )\equiv -\frac 1\pi J\left( \frac \omega \Delta \right) 
$, $s_2(\omega )\equiv -\frac 1\pi J\left( \frac \omega \Delta \right) \frac %
\omega \Delta $, $s_3(\omega )\equiv -\frac 1\pi I\left( \frac \omega \Delta
\right) \frac \omega \Delta $, $s_4(\omega )\equiv -\frac 1\pi J\left( \frac %
\omega \Delta \right) \frac{\omega ^2}{\Delta ^2}$, and $\omega _0\equiv
\omega $, $\omega _1\equiv \omega -E_{\uparrow }-E_{\downarrow }-U$, $\omega
_2\equiv \omega +E_{\uparrow }+E_{\downarrow }+U$, $\omega _3\equiv \omega
-E_{\uparrow }+E_{\downarrow }$. The dimensionless functions $I$, $J$, $Q$
are defined as 
\begin{eqnarray}
I(x) &\equiv &\left\{ 
\begin{array}{ll}
\frac{\arcsin x}{x\sqrt{1-x^2}}\;, & \left| x\right| <1 \\ 
\frac{\text{i}\pi }{2x\sqrt{x^2-1}}-\frac 1{\left| x\right| \sqrt{x^2-1}}\ln
\left( \left| x\right| +\sqrt{x^2-1}\right) \;, & \left| x\right| >1
\end{array}
\right. \; \\
J(x) &\equiv &\left\{ 
\begin{array}{ll}
\frac \pi 2\frac 1{\sqrt{1-x^2}}\;, & \left| x\right| <1 \\ 
\frac \pi 2\frac{\text{i}}{\sqrt{x^2-1}}sign(x)\;, & \left| x\right| >1
\end{array}
\right. \; \\
Q(x,y) &\equiv &-\frac{\text{i}}2\left( \frac 1{e^x+1}+\frac 1{e^y+1}\right)
+\frac 1{2\pi }%
\mathop{\rm Re}%
\left[ \psi \left( \frac 12+\frac{ix}{2\pi }\right) -\psi \left( \frac 12+%
\frac{iy}{2\pi }\right) \right] ,
\end{eqnarray}
with $\psi $ being the digamma function.

As for ${\bf G}^{<}$, we invoke the stationary condition 
\begin{equation}
\left( 
\begin{array}{cc}
\langle \frac d{dt}c_{\uparrow }^{\dagger }c_{\uparrow }\rangle & \langle 
\frac d{dt}c_{\downarrow }c_{\uparrow }\rangle \\ 
\langle \frac d{dt}c_{\uparrow }^{\dagger }c_{\downarrow }^{\dagger }\rangle
& \langle \frac d{dt}c_{\downarrow }c_{\downarrow }^{\dagger }\rangle
\end{array}
\right) =0\;.
\end{equation}
Using Hensenburg equation, one can derive 
\begin{equation}
\int \frac{d\omega }{2\pi }\left[ {\bf G}^{<}{\bf \Sigma }_0^a{\bf +G}^r{\bf %
\Sigma }_0^{<}{\bf -\Sigma }_0^r{\bf G}^{<}{\bf -\Sigma }_0^{<}{\bf G}^a{\bf %
+}\frac{E_{\uparrow \downarrow }}2\left( {\bf G}^{<}{\bf \sigma }_z{\bf %
-\sigma }_z{\bf G}^{<}\right) \right] =0\;,
\end{equation}
where $E_{\uparrow \downarrow }\equiv E_{\uparrow }+E_{\downarrow }+U$ and $%
{\bf \sigma }_z$ is the 3rd Pauli matrix. We remove the integral over energy
and write down the equation for ${\bf G}^{<}$ as 
\begin{equation}
\left( {\bf \Sigma }_0^r+\frac 12E_{\uparrow \downarrow }{\bf \sigma }%
_z\right) {\bf G}^{<}-{\bf G}^{<}\left( {\bf \Sigma }_0^a+\frac 12%
E_{\uparrow \downarrow }{\bf \sigma }_z\right) ={\bf G}^r{\bf \Sigma }_0^{<}%
{\bf -\Sigma }_0^{<}{\bf G}^a\;,
\end{equation}
which can be interpreted as ``detailed'' stationary condition, i.e., the
observables in the energy interval $\omega $ and $\omega +d\omega $ are
time-invariant in the steady state. We note that Eq.(16) is exact while
Eq.(17) is an approximation which neglects high order fluctuations in the
presence of strong correlation (for details see \cite{Gd}). Nevertheless,
this approximation is sufficient for the study of ARISP which works in the
Coulomb blockade regime. In addition, Eq.(17) becomes exact in the $%
U\rightarrow 0$ limit. For comparison, we also try the Keldysh equation $%
{\bf G}^{<}{\bf =G}^r{\bf \Sigma }^{<}{\bf G}^a$, and employ the commonly
used Ng's ansatz for ${\bf \Sigma }^{<}$ \cite{Ng,EOM2}. Although the
calculated results are qualitatively agree with each other, the convergence
is much poorer in the latter approximation scheme. To sum up, Eq.(5) for $%
{\bf G}^r$ and Eq.(17) for ${\bf G}^{<}$ will be applied to the numerical
study of ARISP.

Before presenting the numerical results, we qualitatively analyze the
physics in the problem. As seen in the non-interacting case, spin
polarization occurs on the AR\ resonance. The conditions for AR resonance
are: (1) QD is occupied with even number electrons and (2) a pair of
electrons can transfer freely between QD and S. In the noninteracting case,
these conditions amount to $\mu _L>\mu _R$ and $E_0=\mu _R$ for AR dip, or $%
\mu _L<\mu _R$ and $E_0=\mu _R$ for AR peak. In the presence of Coulomb
interaction, the occupation number of QD is quantized, and AR occurs in the
even occupation region. Notice that the occupation of QD is mainly
determined by F, let us cut off the S-QD coupling and consider the electron
filling in F-QD system. QD has four occupation configurations: $\left|
0\right\rangle $, $\left| \uparrow \right\rangle $, $\left| \downarrow
\right\rangle $ and $\left| \uparrow \downarrow \right\rangle $, with energy 
$0$, $E_{\uparrow }-\mu _F$, $E_{\downarrow }-\mu _F$ and $E_{\uparrow
}+E_{\downarrow }+U-2\mu _F$, respectively. It is energy favorable that QD
is empty when $E_0-\mu _F>0$, singly occupied when $-U<E_0-\mu _F<0$, and
doubly occupied when $E_0-\mu _F<-U$. On the other hand, AR resonance
requires energy degeneracy when moving two electrons from QD to S, so that
Coulomb blockade is lifted. It is straight forward to write down the
equality, $E_{\uparrow }+E_{\downarrow }+U=2\mu _S$, meaning that a spin$%
\uparrow $ and a spin$\downarrow $ electron of QD enter S and form a Cooper
pair at the chemical potential $\mu _S$. Therefore the conditions for AR
resonance are: $E_0-\mu _S=-\frac U2$ and $E_0-\mu _F<-U$ for AR dip, or $%
E_0-\mu _S=-\frac U2$ and $E_0-\mu _F>0$ for AR peak. The key point is that
electron filling is linked to $\mu _F$ while AR\ resonance is linked to $\mu
_S$, and $\mu _F\neq \mu _S$ in nonequilibrium.

Fig.3a shows the curve of $\langle n_\sigma \rangle $ vs $E_0$ for $p=0.5$.
As expected, a spin polarized AR\ dip is superposed on a step-like pattern.
The steps from $0$ to $\frac 12$ and $\frac 12$ to $1$ are located around $%
E_0-\mu _F=0$ and $E_0-\mu _F=-U$, while AR dip located around $E_0-\mu _S=-%
\frac U2$, which are in coincidence with the above discussion. $\langle
n_{\uparrow }\rangle $ and $\langle n_{\downarrow }\rangle $ are nearly
equal off the AR resonance. This is because the weak F-QD coupling can only
induce small spin-dependent renormalization and broadening to $E_{\uparrow }$
and $E_{\downarrow }$. On the contrast, $\langle n_{\uparrow }\rangle $ and $%
\langle n_{\downarrow }\rangle $ are dramatically different on the AR
resonance, which relies on the ratio of $\Gamma _{F\uparrow }$ to $\Gamma
_{F\downarrow }$ rather than their magnitudes. The up-right inset shows the
detailed lineshape of AR dip for several $p$. Comparing with Fig.2a, one can
see that Eq.(4) derived in the non-interacting case can also describe the
ARISP in the presence of Coulomb interaction. In fact, Coulomb interaction
plays the role of quantizing the electron number of QD. Coulomb blockade is
lifted for the conditions discussed above, under which AR occurs as if
through a noninteracting QD. Next, we discuss the choice of charging energy $%
U$, which is determined by the size of QD. Fig.3b and Fig.3c shows the
curves of $\langle n_\sigma \rangle $ vs $U$ on the AR resonance ($E_0-\mu
_S=-\frac U2$). Fig.3b is for the case of $V<\Delta $, in which single
particle process is forbidden due to superconducting gap and AR process
dominates. One can see in the plot that the spin polarization is almost
independent on $U$ when $U<2V$, but gradually suppressed when $U>2V$. The
reason is as follows: on the AR resonance, QD favorites double occupation
when $U<2V$ and single occupation when $U>2V$. For double occupation, as
shown in Eq.(4), ARISP is determined by the ratio $p$ and $r$, and
independent on other parameters. For single occupation, the effective S-QD\
coupling is greatly suppressed, and therefore ARISP vanishes. One tends to
think that increasing the bias voltage $V$ may help to overcome the charging
energy $U$. This is true as long as $V<\Delta $. When $V>\Delta $, however,
both single particle process and AR process are allowed, and the situation
is more complicated. Fig.3c shows $\langle n_\sigma \rangle $ vs $U$ for $%
V>\Delta $. It turns out that the spin polarization is still independent on $%
U$ when $U<2\Delta $, but has a maximum at $U=2\Delta $, suppressed when $%
2\Delta <U<2V$, and gradually vanishes when $U>2V$. The suppression in the
range of $2\Delta <U<2V$ can be attributed to the onset of single particle
process. We note that EOM solution is valid for relatively small $U$, and
the suppression of ARISP is probably underestimated when $U>2\Delta $.
Nevertheless, the maximum at $U=2\Delta $ is reliable, which is related to
the singularity in the density of states of S electrode.

In conclusion, we have proposed an efficient mechanism for writing spin in a
QD which is based on the ARISP effect in F-QD-S system. The scheme has the
advantages that the magnetization of F is not required to be strong and the
sign of spin polarization can be controlled in fully electric manner.
Calculation shows that the optimal conditions for ARISP are: $\Gamma _F\ll
\Gamma _S$, $U=2\Delta $, $E_0-\mu _S=-\frac U2$, and $\left| V\right| >U/2$%
. The properties of ARISP can be described by Eq.(4). In practice, the
resonant level $E_0$ can be tuned by gate voltage, and AR\ resonance can be
monitored by the small tunnel current between F and S electrodes.. In the
context of intensive research and impressive progress in F/2DEG, S/2DEG, and
F/S hybrid structures, the proposed F-QD-S system should be feasible with
up-to-date nano-technology. We are looking forward to hearing the relevant
experimental response.

This project was supported by NSFC\ under Grants No. 10074001 and No.
90103027. T. H. Lin would also like to thank the support from the Visiting
Scholar Foundation of State Key Laboratory for Mesoscopic Physics in Peking
University.

\smallskip $^{*}$ To whom correspondence should be addressed.


\newpage

\section*{Figure Captions}

\begin{itemize}
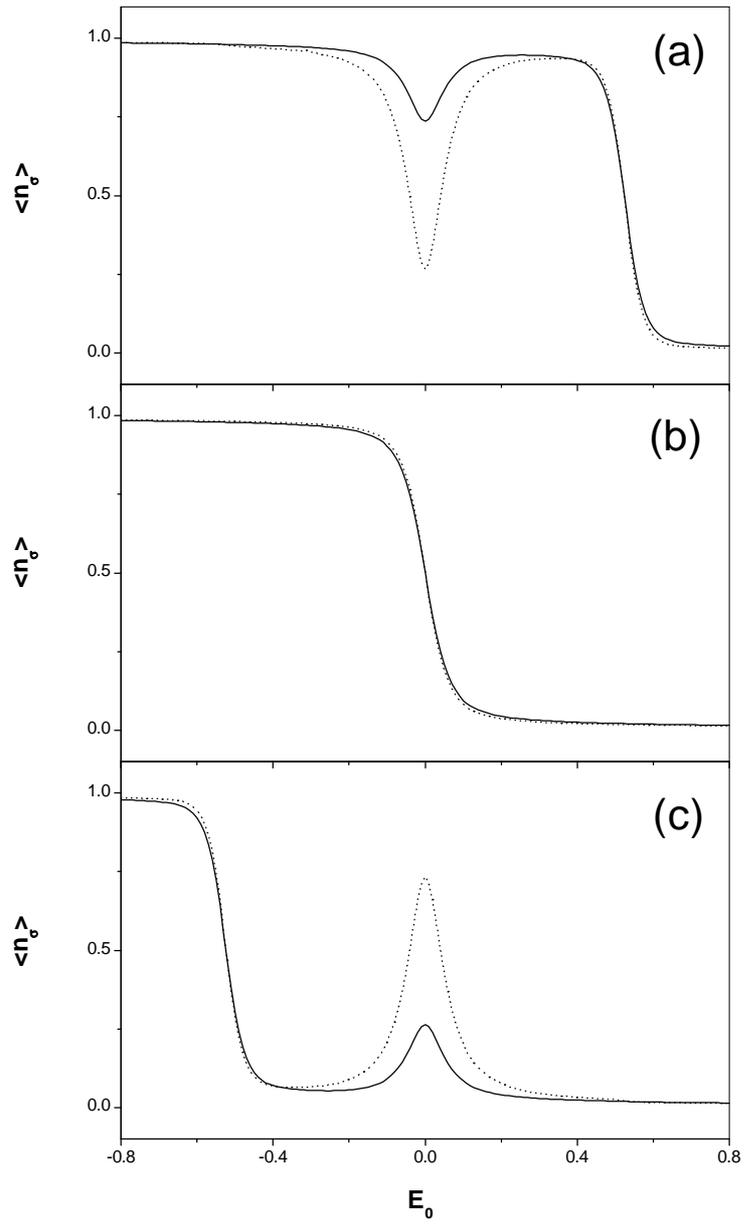
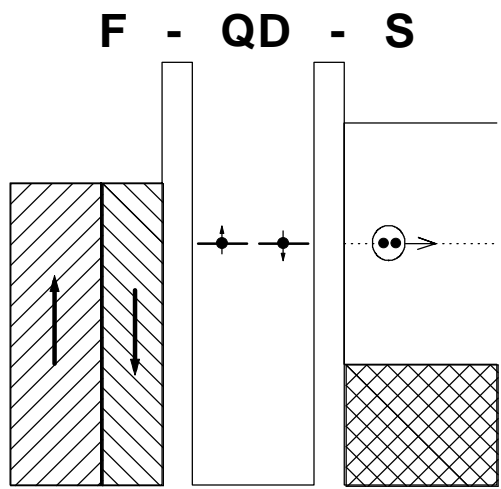
\item[{\bf Fig. 1}]  The occupation number $\langle n_{\uparrow }\rangle $
(solid) and $\langle n_{\downarrow }\rangle $ (dotted) vs the resonant level 
$E_0$ in a non-interacting F-QD-S system. The superconducting gap $\Delta =1$
is set as energy unit, the bias voltage $V=0.5$, $0$, and $-0.5$ for (a),
(b), and (c), respectively. Other parameters are: $\Gamma _F=0.01$, $\Gamma
_S=0.1$, $k_BT=0.02$, and $p=0.5$. The diagram in the bottom schematically
shows the AR resonance in non-interacting F-QD-S system.

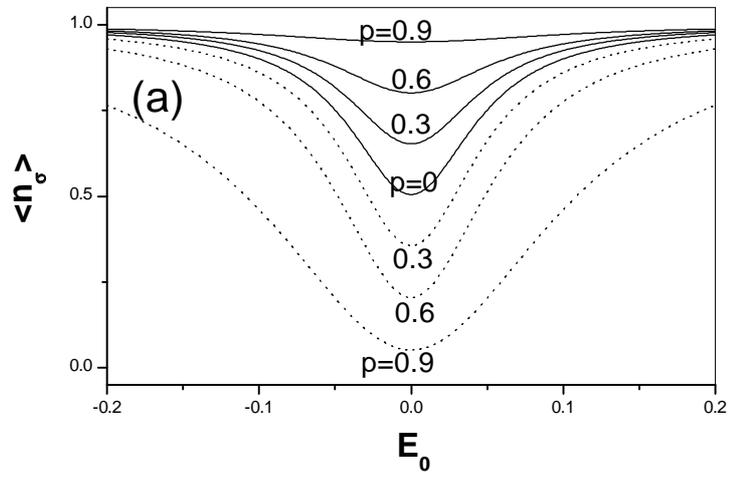
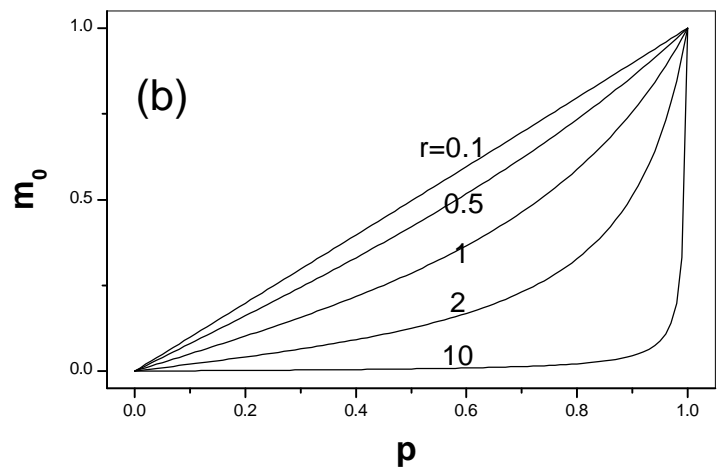
\item[{\bf Fig. 2}]  Detailed analysis of the Andreev dip in Fig.1a: (a) $%
\langle n_{\uparrow }\rangle $ (solid) and $\langle n_{\downarrow }\rangle $
(dotted) vs $E_0$ for different magnetization $p$ of F electrode. (b) $%
m_0\equiv \left( \langle n_{\uparrow }\rangle -\langle n_{\downarrow
}\rangle \right) _{E_0=0}$ vs $p$ for different coupling strength ratio $r$.

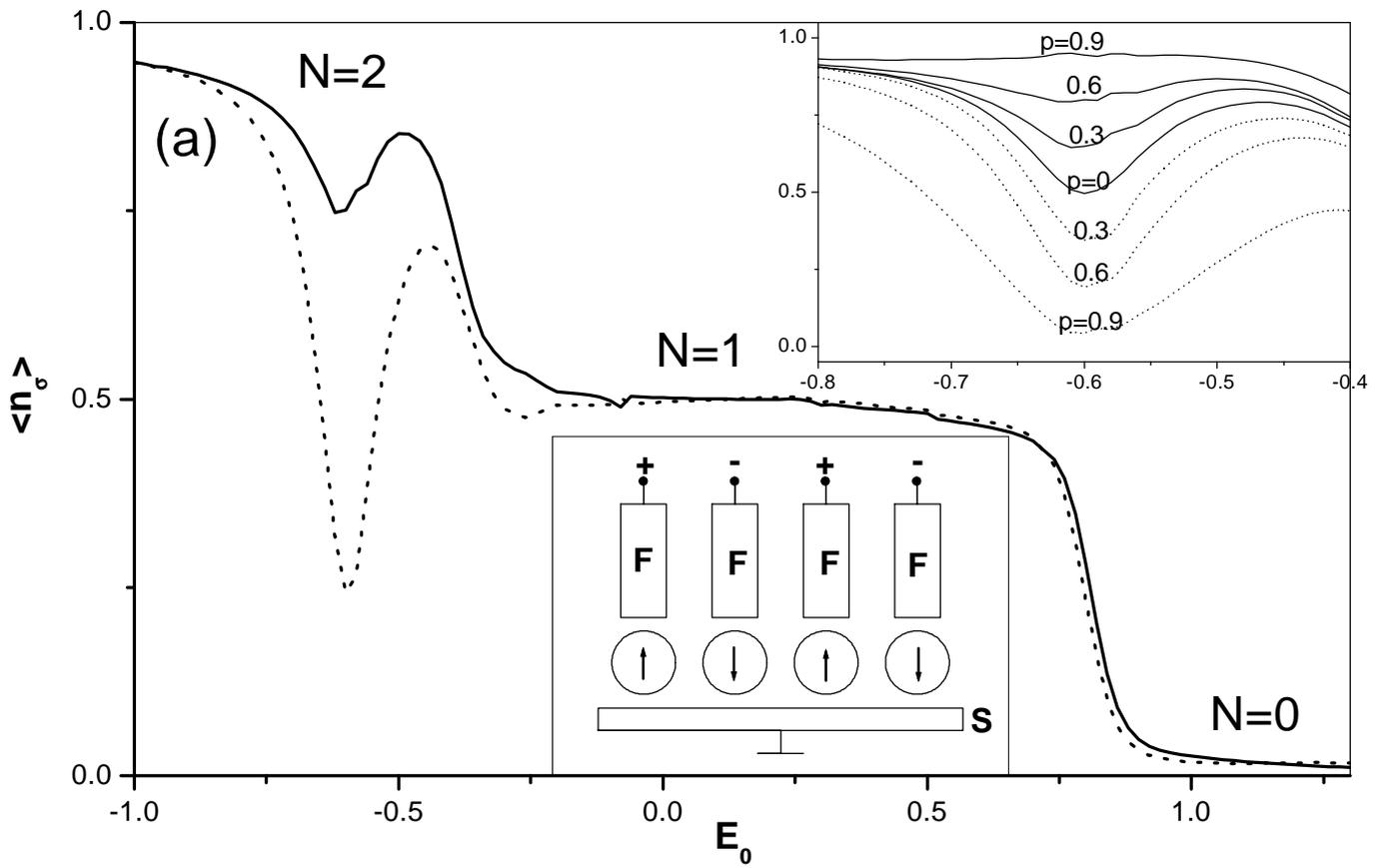
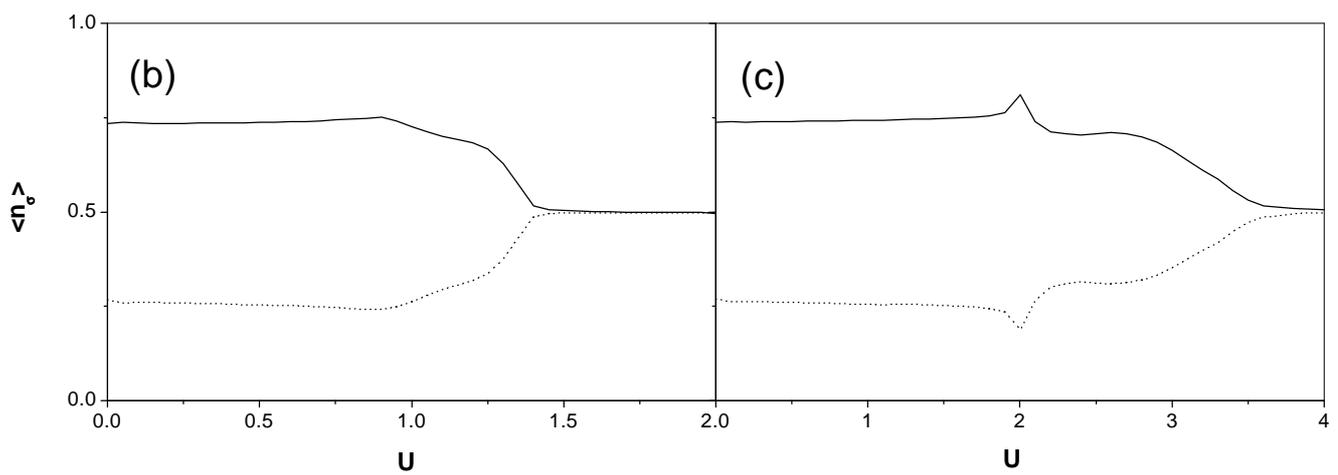
\item[{\bf Fig. 3}]  (a) The occupation number $\langle n_{\uparrow }\rangle 
$ (solid) and $\langle n_{\downarrow }\rangle $ (dotted) vs the resonant
level $E_0$ in an interacting F-QD-S system. The superconducting gap $\Delta
=1$ is set as energy unit, the bias voltage $V=0.75$, and the charging
energy $U=1.2$. Other parameters are the same as Fig.1. The up-right inset
shows the details of the Andreev dip around $E_0-\mu _S=-\frac U2$. The
middle inset illustrates the possibility of using ARISP effect to manipulate
the spin states of a QD array. (b) and (c) show the curves of $\langle
n_{\uparrow }\rangle $ (solid) and $\langle n_{\downarrow }\rangle $
(dotted) at $E_0-\mu _S=-\frac U2$ vs the charging energy $U$, with the bias
voltage $V=0.5$ and $V=1.5$, respectively.
\end{itemize}


\begin{references}
\bibitem{spin1}  For a recent review, see chapter 8 in {\it Semiconductor
Spintronics and Quantum Computation}, B. Burkard and D. Loss, and references
therein.

\bibitem{spin2}  D. D. Awschalom and J. M. Kikkawa, Phys. Today {\bf 52,} 33
(1999).

\bibitem{spin3}  A. V. Khaetskii and Y. V. Nazarov, Phys. Rev. B {\bf 61,}
12639 (2000); {\it ibid}, {\bf 64,} 125316 (2001).

\bibitem{spin4}  P. Recher, E. V. Sukhorukov, and D. Loss, Phys. Rev. Lett. 
{\bf 85,} 1962 (2000).

\bibitem{spin5}  J. Fransson, O. Eriksson, and I. Sandalov, Phys. Rev. Lett. 
{\bf 88,} 226601 (2002).

\bibitem{AR1}  C. W. J. Beenakker, Phys. Rev. B {\bf 46, }12841 (1992).

\bibitem{AR2}  N. R. Clauphton, M. Leadbeater, and C. J. Lambert, J. Phys.:
Condens. Matter {\bf 7, }8757 (1995).

\bibitem{AR3}  R. Fazio and R. Raimondi, Phys. Rev. Lett. {\bf 80,} 2913
(1998).

\bibitem{AR4}  Q. -f. Sun, J. Wang, and T. -h. Lin, Phys. Rev. B {\bf 59, }%
3831 (1999).

\bibitem{remark1}  We choose $e=1$ so that voltage has the energy scale, set 
$\mu _S=0$ and $\mu _F=V$ for convenience.

\bibitem{Andreev}  A. F. Andreev, Zh. Eksp. Teor. Fiz. {\bf 46, }1823 (1964)
[Sov. Phys. JETP {\bf 19, }1228 (1964)].

\bibitem{EOM1}  Y. Meir, N. S. Wingreen, and P. A. Lee, Phys. Rev. Lett. 
{\bf 66, }3048 (1991).

\bibitem{EOM2}  Q. -f. Sun, H. Guo, and T. -h. Lin, Phys. Rev. Lett. {\bf %
87, }176601 (2001).

\bibitem{remark2}  We further suppose that $k_BT\ll \Delta $ in the
evaluation of ${\bf \Sigma }_2^r$.

\bibitem{Ng}  T. -K. Ng, Phys. Rev. Lett. {\bf 76, }487 (1996).

\bibitem{Gd}  Q. -f. Sun, H. Guo, J. Wang, cond-mat/0212157.
\end{references}
\end{document}